\documentclass[%
reprint,
superscriptaddress,
 amsmath,amssymb,
pra,
floatfix,
]{revtex4-2}

\usepackage{xcolor} 
\usepackage{graphicx}
\usepackage{dcolumn}
\usepackage{bm}
\usepackage{booktabs}
\usepackage[english]{babel}
\usepackage[autostyle=true]{csquotes}

\usepackage{hyperref}
\usepackage[mathlines]{lineno}

\begin{document}

\preprint{APS/123-QED}

\title{Wireless millikelvin interconnects for superconducting quantum hardware}

\author{Kristopher Barr}
\altaffiliation{These authors contributed equally to this work.}
\affiliation{
Department of Physics, SUPA, University of Strathclyde, Glasgow G4 0NG, United Kingdom}
\author{Mingyan Zhong}%
\altaffiliation{These authors contributed equally to this work.}
\affiliation{
James Watt School of Engineering, University of Glasgow, Glasgow G12 8QQ, United Kingdom}
\author{Euan Parry}%
\altaffiliation{These authors contributed equally to this work.}
\affiliation{
Department of Physics, SUPA, University of Strathclyde, Glasgow G4 0NG, United Kingdom}
\author{Manoj Stanley}%
\affiliation{National Physical Laboratory, Hampton Road, Teddington TW11 0LW, United Kingdom}
\author{Qusay Al-Taai}%
\affiliation{
James Watt School of Engineering, University of Glasgow, Glasgow G12 8QQ, United Kingdom}
\author{Paniz Foshat}%
\affiliation{
James Watt School of Engineering, University of Glasgow, Glasgow G12 8QQ, United Kingdom}
\author{Kaveh Delfanazari}%
\affiliation{
James Watt School of Engineering, University of Glasgow, Glasgow G12 8QQ, United Kingdom}
\author{Martin Weides}%
\affiliation{
James Watt School of Engineering, University of Glasgow, Glasgow G12 8QQ, United Kingdom}
\author{Nick M. Ridler}%
\affiliation{National Physical Laboratory, Hampton Road, Teddington TW11 0LW, United Kingdom}
\author{Chong Li}%
\email{Email: chong.li@glasgow.ac.uk}
\affiliation{
James Watt School of Engineering, University of Glasgow, Glasgow G12 8QQ, United Kingdom}
\author{Alessandro Rossi}
\email{Email: alessandro.rossi@strath.ac.uk}
\affiliation{
Department of Physics, SUPA, University of Strathclyde, Glasgow G4 0NG, United Kingdom}
\affiliation{National Physical Laboratory, Hampton Road, Teddington TW11 0LW, United Kingdom}

\date{\today}

\begin{abstract}
Scalable quantum computing is limited by the dense network of electrical interconnects linking cryogenic quantum processors to room-temperature control electronics. To overcome this bottleneck, considerable effort has focused on cryogenic CMOS electronics and microwave-to-optical transduction, aiming to reduce wiring complexity and thermal loading. Wireless interconnects have recently emerged as a promising complementary approach, yet their compatibility with superconducting quantum hardware remains largely unexplored. Here, we demonstrate the wireless excitation of a superconducting microwave resonator of the type routinely employed for qubit readout, operating at millikelvin temperatures inside a dilution refrigerator. By directly comparing wired and wireless operation within the same cryogenic environment, we show that wireless coupling preserves the intrinsic resonator response while revealing parasitic electromagnetic pathways arising from stray radiation within the cryostat enclosure. These results establish a framework for the co-design of wireless interconnects, cryogenic packaging and superconducting quantum hardware.

\end{abstract}

\maketitle


\section{Introduction\label{sec:intro}}
Quantum computers capable of delivering practical advantage are expected to require thousands of logical qubits and ultimately millions of physical qubits~\cite{Campbell2017}. Achieving this scale presents formidable engineering challenges, as current quantum processors rely on dense bundles of interconnects linking cryogenic quantum hardware to room-temperature electronics. In dilution refrigerators, these interconnects occupy substantial physical volume, introduce significant thermal loads, and increasingly limit system scalability~\cite{krinnerEngineeringCryogenicSetups2019}. Overcoming the interconnect bottleneck has therefore emerged as a central challenge on the path towards fault-tolerant quantum computing.

Several approaches are currently being pursued to address this challenge~\cite{awsch-rev, brennanClassicalInterfacesControlling2025}. One strategy consists of integrating classical electronics closer to the quantum processor through cryogenic electronics (cryo-CMOS), thereby reducing the number of room-temperature connections while enabling local control, multiplexing, and signal processing~\cite{Xue2021,charbon-rev2, acharya_multiplexed_2023}. A complementary approach relies on microwave-to-optical conversion and photonic interconnects, which offer low thermal conductivity and the prospect of long-distance communication~\cite{Lecocq2021, arnold_all-optical_2025}. At present, cryo-CMOS architectures remain constrained by local wiring density and power dissipation~\cite{zouPowerDeliveryCryogenic2025}, whereas photonic approaches face challenges associated with conversion efficiency and device integration~\cite{lamb-rev}.

\begin{figure*}[!htbp]
    \includegraphics{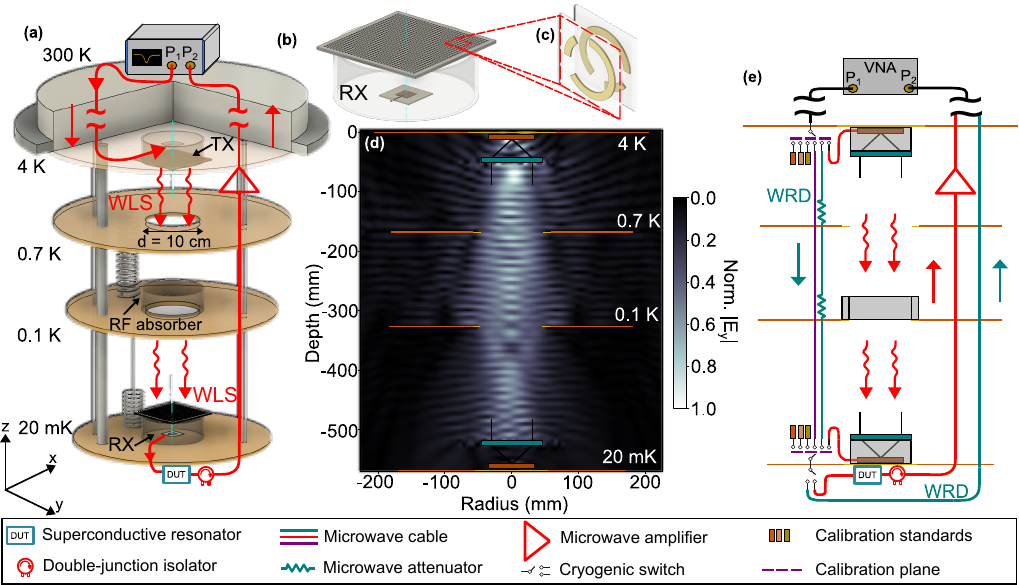}
    \caption{\label{fig:design} (a) Schematic diagram of the WLS excitation link inside a dilution refrigerator ($50~$K stage is omitted for clarity). Shaded cylinders indicate radiation absorbers used to attenuate reverberant signals. (b) RX module composed of a patch antenna and a flat metalens, the absorber attenuates the signal that is not transmitted through the metalens. (c) Zoom-in detail of the DSRR metalens pattern (front and back sides) flipped across the y = -x axis. (d) Simulations of the y-polarized electric field, $E_\text{y}$, propagating within the cryostat chamber. The TX metalens collimate the signal through the apertures, whereas the RX metalens refocuses it on the MXC antenna. (e) Schematic diagram of the full set-up with both WLS (red) and WRD (green) channels, cryogenic attenuators and switches, and calibration standards.}
\end{figure*}
 
Wireless interconnects have recently emerged as a promising alternative route towards scalable quantum computing architectures~\cite{alarconScalableMultichipQuantum2023}. By replacing physical transmission lines with free-space electromagnetic links, wireless approaches offer the prospect of reducing thermal loading, alleviating routing congestion, and enabling reconfigurable communication pathways between cryogenic subsystems~\cite{zouPowerDeliveryCryogenic2025}. Consequently, wireless architectures have recently been proposed, ranging from gigahertz-frequency intra-cryostat protocols~\cite{bandara28GHzWireless2025} to terahertz cryogenic experimental interconnects~\cite{wangWirelessTerahertzCryogenic2025}. These studies have preliminarily explored the feasibility of wireless communication in cryogenic environments, including antenna design, channel characterization, and information-transfer efficiency. The interaction between wireless electromagnetic fields and superconducting quantum hardware has received comparatively little attention. Quantum processors are embedded within highly reflective metallic enclosures that support complex electromagnetic environments, potentially giving rise to unintended radiation pathways. Understanding these effects is essential before wireless technologies can be realistically deployed in large-scale quantum computing systems.\\\indent
Here, we experimentally demonstrate the wireless excitation of a superconducting resonator similar to those routinely employed for qubit readout, operating at millikelvin temperatures inside a dilution refrigerator. By directly comparing wired and wireless operation within the same cryogenic environment, we show that wireless coupling preserves the intrinsic resonator response. We further reveal electromagnetic interactions arising from stray radiation within the highly reflective cryostat cavity. We show that, although RF absorbers effectively suppress reverberation effects, residual parasitic coupling to the device persists through unintended propagation pathways. These findings establish practical design guidelines for wireless millikelvin interconnects and highlight the need for the co-design of wireless links, cryogenic packaging, and superconducting quantum hardware in future scalable quantum computing systems.
\section{Wireless \& Wired Links\label{sec:main}}
To establish a wireless (WLS) cryogenic link inside our dilution refrigerator, we designed and manufactured a transmitter/receiver module (TX/RX) tailored to beam microwave radiation through a line-of-sight (LOS) opening. Figure~\ref{fig:design}(a) shows the TX mounted underneath the $4~$K stage and the RX mounted on the mixing chamber (MXC) stage.  Prior to cryogenic operations, RX and TX are carefully aligned across an aperture (diameter $d=10~$cm) machined into the still and cold stages of the refrigerator. Considerations around the heat load introduced by the LOS arrangement can be found in  Appendix~\ref{supp:heatload}. As shown in Fig.~\ref{fig:design}(b), the TX module consists of a planar patch antenna and a flat metalens designed to collimate the radiation beam and is designed to operate at frequencies around $10.6~$GHz (see Appendix~\ref{supp:efficiency}). The RX module, operating at $T_{\text{mxc}} = 20$ mK, incorporates a metalens that refocuses the transmitted beam onto the receiving antenna. As depicted in Fig.~\ref{fig:design}(c), the lens architecture is composed of two double split-ring resonators (DSRRs) patterned on the front and back of the metalens (see Methods and Appendix \ref{supp:DSRRs}). At the design stage, such transmission properties were modeled using commercial full-wave electromagnetic simulation software, see Fig.~\ref{fig:design}(d) and Appendix~\ref{supp:DSRRs}. Radiation absorbers (Eccosorb, Laird) were installed around the cold stage aperture, as well as around TX and RX modules to reduce reverberations and multi-path reflections. Note that in the final leg of the transmission path, the RX is connected onto our testbed superconductive resonator device (DUT) via short coaxial lines running across the MXC stage.\\\indent
\begin{figure*}[!htbp]
    \includegraphics{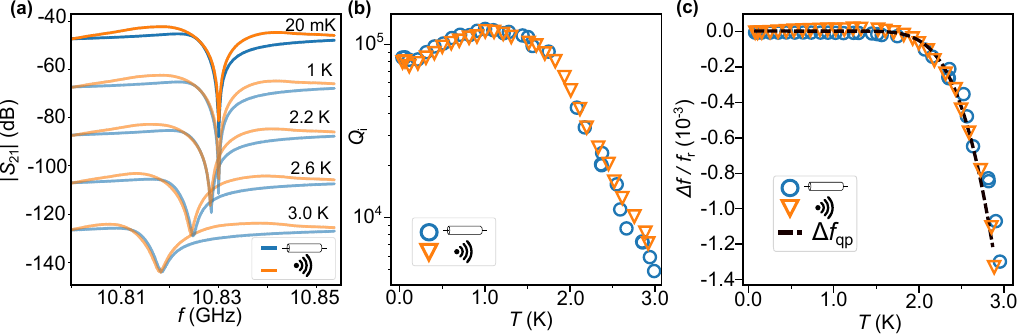}
    \caption{\label{fig:CPW} (a) Transmission coefficient as a function of frequency for different MXC temperatures in WLS (orange) and WRD (blue) excitation mode. Each pair of isothermal traces has been shifted vertically by multiples of 20 dB for clarity. (b) $Q_{\text{i}}$ as a function of temperature extracted in the WLS (triangle) and WRD (circle) mode. (c) Fractional change in resonant frequency as a function of temperature in WLS (triangle) and WRD (circle) mode. The dashed line indicates the expected temperature dependence contribution arising from quasi-particles excitation, $\Delta f_{\text{qp}}$ (see also Appendix \ref{supp:resonator}).}
\end{figure*}
To benchmark the performance of the wireless link against a well known reference, we also implemented a fully wired (WRD) channel, as shown in Fig.~\ref{fig:design}(e). Besides functioning as a validation  experiment, the wired link enables cryogenic calibration of the Vector Network Analyzer (VNA), as discussed in Methods. The wired channel consists of UT085 stainless steel coaxial lines with 40 dB total attenuation. Seamless alternation between wireless and wired configurations could be obtained during the experiments through the operation of three cryogenic switches, one located at the $4~$K stage and two at the MXC stage, shown in Fig.~\ref{fig:design}(e). Besides comparing the excitation of the DUT in either wired or wireless mode, an appropriate choice of switch configurations enabled us to bypass the DUT altogether, with the aim of characterizing the wireless and/or wired assembly in isolation.\\\indent
It is important to note that although the excitation link could be selected to be wired or wireless at will, the output path for the signal was always wired in our experiments. This consisted of a standard double junction magnetic isolator at the MXC stage, superconductive coaxial lines and a $4~$K amplification stage.

\section{Wireless Resonator Excitation}\label{sub: wireless excitation}

To assess whether wireless excitation preserved the intrinsic properties of the resonator, we benchmarked its resonant frequency ($f_\textup{r}$), internal quality factor ($Q_\textup{i}$), and temperature-dependent fractional frequency shift ($\Delta f / {f_r}$) against wired measurements. These metrics were measured over the temperature range $20~$mK$\leq T_{\text{mxc}} \leq 3.0~$K to probe consistency across fundamental dissipation mechanisms. An approximate resonator input power of -116 dBm was maintained in both configurations, corresponding to operation in the few-photon regime.\\\indent
Figure~\ref{fig:CPW}(a) shows the evolution of the transmission resonance at five different temperatures of operation for both the wired and wireless excitation modes. It shows broad agreement between the resonance lineshapes over the investigated temperature range. However, the principal difference is the absence of an asymmetric shoulder under wireless excitation. This suggests that wireless excitation modifies the coupling of the incident field to the resonator-feedline system. In fact, such type of asymmetry in the wired resonance is usually attributed to Fano interference arising from background transmission pathways~\cite{riegerFanoInterferenceMicrowave2023} and impedance mismatches. Its suppression under wireless excitation suggests that these interference pathways are substantially altered. Circle-fitting~\cite{probstEfficientRobustAnalysis2015a} enabled extraction of the quality factors and resonant frequency shifts as functions of temperature, see Fig.~\ref{fig:CPW}(b)(c). \\\indent The close agreement in the temperature dependence demonstrates that wireless excitation properly reproduces the intrinsic resonator response as in the traditional wired mode. In particular, well-known dissipation mechanisms remain at play no matter which excitation is selected. Both excitation schemes exhibit the characteristic increase in $Q_\textup{i}$ at low temperature associated with reduced two-level-system loss, followed by the monotonic degradation expected from thermally generated quasiparticles. Likewise, the fractional frequency shift follows the expected quasiparticle-induced evolution throughout the investigated temperature range~\cite{mcraeMaterialsLossMeasurements2020, alexanderPowerTemperatureDependent2025a}.\\\indent
These results demonstrate that wireless excitation preserves the intrinsic superconducting response of the resonator, providing strong evidence that free-space microwave delivery is compatible with the operation of superconducting devices used for quantum readout.

\section{Stray Wireless Losses}\label{sec:parasitic_wless}
\begin{figure}[t]
    \includegraphics{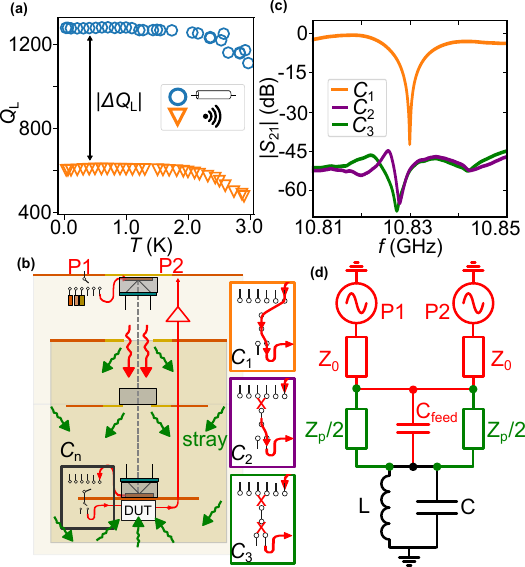}
    \caption{\label{fig:parasitic} 
    (a) $Q_{\text{L}}$ as a function of temperature for WLS (triangles) and WRD (circles) configurations.
    (b) Illustration of pathways of stray radiation (green arrows) and intended line-of-sight path (red arrows) within the cryostat. The switches enclosed in the black rectangle at the MXC stage are used in three alternative configurations $[C_1, C_2, C_3]$ shown on the right and color coded according to the measured transmission signal. 
    (c) DUT transmission coefficient as a function of frequency for three configurations of the cryogenic switches: intended WLS link $C_1$ (orange), single broken link $C_2$ (purple) and double broken link $C_3$ (green).
    (d) Equivalent circuit model including the intended WLS path (red), the additional stray path (green) and the resonator lumped elements (black). 
    }
    \end{figure}

\begin{figure}[t]
        \includegraphics{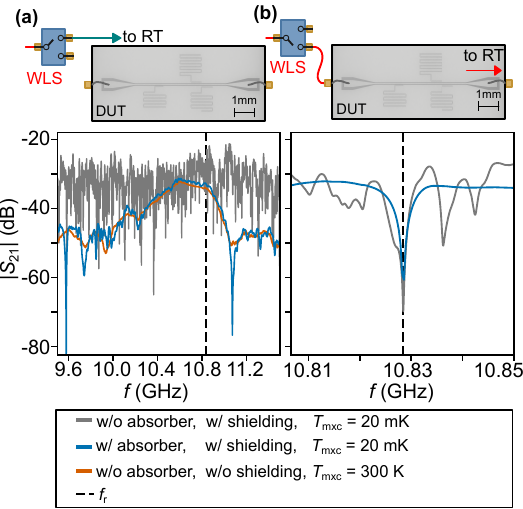}
        \caption{\label{fig:Snn} (a) Top: schematic diagram of the meander resonator device used as DUT and routing that bypasses it. Bottom: Calibrated $|S_{21}|$ measurements at 20 mK as a function of frequency with (blue) and without (gray) RF absorbers in the refrigerator. Room temperature unshielded data is shown in orange. The vertical dashed line highlights the frequency $f_\text{r}$ used for characterizing the DUT response. (b) Top: schematic diagram of the meander resonator device used as DUT and routing through it. Bottom: $|S_{21}|$ measurements as a function of frequency with (blue) and without (gray) RF absorbers.}
    \end{figure}

Despite the broad agreement in intrinsic parameters, the two approaches to excite the resonator have inherent differences that we discuss next. In particular,  a stark discrepancy emerges in the values of the loaded quality factor, $Q_\textup{L}$. Figure~\ref{fig:parasitic}(a) shows that the wireless configuration exhibits a systematically lower value, with an offset $|\Delta Q_\textup{L}| = |Q_\textup{L}^{\text{WLS}} - Q_\textup{L}^{\text{WRD}}| \approx 650$ across the entire temperature range. Because both configurations display an analogous temperature dependence, linked to thermally generated quasiparticles, this constant offset indicates an additional loss channel unique to the wireless link.\\\indent
To study this behavior, we consider two concurrent propagation pathways inside the cryostat environment, see Fig.~\ref{fig:parasitic}(b). We argue that the intended LOS channel implemented via the TX/RX modules operates in tandem with a secondary stray channel arising from radiation diffracted away from the LOS openings and reverberating within the cryostat through multiple reflections facilitated by the refrigerator's radiation shields. To verify this, we use the two cryogenic switches at the MXC plate to deliberately interrupt the path between the wireless link and the DUT, and check whether it is nonetheless excited by a concurrent mechanism. Specifically, we operate the switches in three possible configurations and show the outcome of transmission measurements for each one in Fig.~\ref{fig:parasitic}(c). In configuration $C_1$, the expected transmission through the intended line-of-sight channel and DUT is observed. Interestingly, when the path from the RX module to the DUT is broken by putting one of the switches into an open state (configuration $C_2$), a residual resonant feature persists with a $\sim$50~dB lower baseline and an asymmetric lineshape. This feature proves the existence of a stray pathway, but it does not rule out that this could be the result of poor switch isolation (nominally -60 dB). However, increasing the isolation of the DUT by putting both switches in the open state (configuration $C_3$), the transmission still yields a comparable baseline and resonator readout, albeit with a markedly different lineshape. The absence of signal degradation when both switches are in the open state allows one to rule out poor isolation as the only source of spurious DUT excitation. Instead, these results suggest that stray radiation propagates within the highly reflective cryostat interior and couples to the DUT in parallel to the intended wireless line-of-sight path.\\\indent
One can represent the two measured excitation pathways with the equivalent circuit model presented in Fig.~\ref{fig:parasitic}(d). The intended wireless link (red), with characteristic impedance $Z_\textup{0}$, couples to the resonator through the feedline--resonator geometric capacitance, $C_{\text{feed}} \sim 15~\mathrm{fF}$. The stray pathway (green) is modeled as a parasitic complex impedance, $Z_\textup{p}$, that forms additional symmetric excitation and return paths between the DUT and the readout line. This is used to represent stray radiation incident on the DUT and on the copper enclosure, making it an active part of the resonator's microwave environment. 
We expect that $Z_\textup{p}$ would have a capacitive component determined by the geometry of the enclosure (see Appendix \ref{supp:Q3D}), as well as an inductive component arising from radiation-induced currents in the enclosure. Such coupling channel between enclosure modes and readout line due to stray excitations would be compatible with the observed reduction in $Q_\textup{L}$. 
\\\indent Further evidence around the presence of stray radiation propagating inside the refrigerator is provided by comparing the characterization of the wireless link with and without the RF absorbers of Fig.~\ref{fig:design}(a). In these tests, we configured the switches in order to route the signals through the wireless link at the input line with a bypass on the DUT at the output line, see Fig.~\ref{fig:design}(e). This enabled us to measure the performance of the wireless link in isolation and find out whether the RF absorbers would be needed to avoid spurious interference effects. In Fig.~\ref{fig:Snn}(a), one can see that the transmission spectrum without the absorber (gray trace) exhibits pronounced fluctuations. These ripples are hallmarks of multi-path cavity reverberations enhanced by the radiation shields acting as cavity reflectors. This interpretation is supported by the fact that the removal of the radiation shields during room temperature tests was sufficient to remove the vast majority of those features (see orange trace). Given that shielding is essential for operating the dilution refrigerator, we resolved to introduce the RF absorbers. Such solution resulted in achieving nearly identical transmission performance between shielded and unshielded configurations as indicated by the agreement between orange and blue traces. The practical benefit of this mitigation on device readout is highlighted in Fig.~\ref{fig:Snn}(b). When routing the wireless path through the DUT without RF absorbers in place (gray), the stray background fluctuations severely distort the signal, rendering reliable resonator parameter extraction impossible. Introducing the absorbers (blue) attenuates this multi-path reverberation, flattens the spectrum, and restores a clean resonance lineshape.\\\indent In summary, these results show that RF absorbers play a critical role in mitigating cavity reverberations within the cryostat environment and their introduction is pivotal to reliable extraction of resonator parameters. However, the absorbers do not constitute a complete solution to stray wireless coupling. Even in their presence, residual radiation continues to propagate through unintended pathways and couples to the DUT, as evidenced by the persistent offset in loaded quality factor.

\section{\label{sec:dis}Conclusion and Outlook}
We have demonstrated wireless excitation of a superconducting resonator operating at millikelvin temperatures inside a dilution refrigerator. By directly comparing wired and wireless operation, we show that free-space microwave delivery preserves the intrinsic electrodynamic response of the device, including its resonant frequency, internal quality factor, and temperature-dependent frequency shift. These results provide strong evidence that wireless links are fundamentally compatible with superconducting hardware used for quantum readout.

At the same time, our measurements reveal important challenges associated with implementing wireless interconnects in cryogenic environments. In particular, stray radiation propagating through the highly reflective interior of the refrigerator introduces additional coupling pathways that are absent in conventional wired architectures. RF absorbers are shown to be highly effective in suppressing cavity reverberations and restoring clean resonance spectra, yet they do not entirely eliminate residual coupling to the device. This observation highlights that controlling reverberation and controlling stray device excitation are distinct engineering problems that must both be addressed in future implementations.

The results presented here establish a first experimental framework for identifying and quantifying these effects. Future wireless cryogenic architectures will likely require a combination of absorptive materials, optimized enclosure geometries, and device packaging strategies~\cite{levenson-falkReviewDesignConcerns2025, huangMicrowavePackageDesign2021} that minimize parasitic coupling between quantum hardware and its electromagnetic environment. Such considerations become increasingly important as wireless approaches evolve towards multi-channel~\cite{alarconScalableMultichipQuantum2023}, multi-chip, or higher-frequency architectures~\cite{bandara28GHzWireless2025}. While further engineering advances will be required, our findings suggest that wireless interconnects constitute a credible route for alleviating the physical volume and thermal-management challenges associated with conventional cryogenic wiring.


\section{\label{sec:meth}Methods}
The DUT is a notch-type $\lambda / 4 $ meander resonator patterned on a 100-nm-thick NbN film on silicon (see Fig.~\ref{fig:Snn}). It was measured at the second harmonic, $f_r = 3\cdot f_0$ with $f_0 \simeq 3.6$ GHz. The chip was housed in a shielded, oxygen-free copper enclosure and thermally anchored to the MXC. Comprehensive fabrication details are provided in Ref.~\cite{foshatQuasiparticleDynamicsNbN2025}.

VNA calibrations employed a short-open-load-reciprocal (SOLR) technique~\cite{stanleyCharacterizingSParametersMicrowave2024, obertoFullTwoPortSParameters2026a} with calibration planes positioned as shown in Fig.~\ref{fig:design}(e). The procedure was carried out with two sets of cryogenic SOL standards and a fully wired (nominally unattenuated) transmission line by relevant choice of switch configurations.\\\indent 
The wireless link is established through dedicated TX and RX modules designed to transfer microwave radiation through the free-space inter-stage apertures of the dilution refrigerator. Each module incorporates a planar feed antenna and a bi-layer metasurface lens~\cite{al-moathinCharacterizationCompactWideband2023}. The feed antenna launches and receives the microwave signal, while the metasurface lens shapes the electromagnetic wavefront to improve transmission efficiency through the apertures. Specifically, the TX lens transforms the spherical wavefront emitted by the antenna into a quasi-plane wave, whereas the RX lens refocuses the transmitted beam onto the receiving antenna. Each metalens comprises a $21 \times 21$ array of unit cells formed by two axially aligned double split-ring resonators patterned on opposite sides of a Rogers 4350B $0.762$-mm-thick substrate and separated by a $1.5$-mm-wide spacer. The TX/RX pair operates through orthogonal polarization conversion: the feed antenna emits an x-polarized microwave signal that is converted into a y-polarized propagating beam by the TX lens and subsequently restored to x-polarization by the RX lens. This polarization-selective architecture reduces radiative crosstalk and suppresses multi-path interference. In addition, the feed antennas provide a high passband-to-stopband ratio that attenuates both low-frequency shot noise and high-frequency blackbody radiation. Further details of the antenna and metalens design are provided in Refs.~\cite{al-moathinCharacterizationCompactWideband2023,rehmankazimWirelessMicrowaveSignal2023} and Appendix~\ref{supp:DSRRs}.


\begin{acknowledgments}
We wish to thank T. Lindstr\"om, A. Zotov and C. McGeough for useful discussions, as well as A. Robbins for technical support. AR acknowledges support from the UKRI Future Leaders Fellowship Scheme (Grant agreement: UKRI1071). CL acknowledges support from EPSRC WiQC project (Grant agreement: EP/X017613/1). MW, CL, NMR and AR acknowledge support from EPSRC project EPIQC (Grant agreement: EP/W032627/1).
\end{acknowledgments}
\section*{Contributions}
K.B., E.P. and M.S. performed the experiments. M.Z. and Q.A.T. designed, fabricated and characterised the RX/TX modules, and performed simulations. P.F. and K.D. designed and fabricated the DUT. K.B. and E.P. carried out data analysis. M.W., N.M.R., C.L. and A.R. devised the experiments and supervised the project. K.B. and A.R. prepared the manuscript with contributions from all authors.

\appendix

\section{\label{supp:heatload}Heat Load}
Thermal loading is a key constraint on the scalability of cryogenic interconnects in dilution refrigerators. Line-of-sight (LOS) apertures between temperature stages provide a pathway for radiative heat transfer, which can impose an unacceptable thermal burden on the coldest stages. Quantifying heat loads produced by a wireless link is therefore essential to evaluate its performance in comparison to a wired link. We estimate the radiative loads introduced by our wireless implementation and examine how these loads scale with aperture diameter and the choice of connected temperature stages.\\\indent
    \begin{figure}[b]
        \includegraphics[scale=1.1]{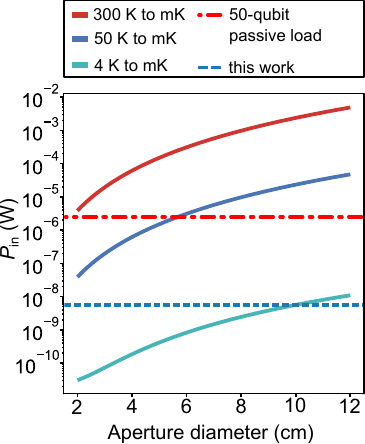}
        \caption{\label{fig:passiveload} Radiative heat load incident on the MXC stage as a function of aperture diameter for radiation propagating through the line of sight from RT stage (red solid trace), PT1 stage (blue solid trace), and PT2 stage (cyan solid trace). The blue dashed line indicates the radiated power expected for our experiments. The red dashed line shows the passive heat load of the WRD 50-qubit set-up reported in Ref.~\cite{krinnerEngineeringCryogenicSetups2019}.}
    \end{figure}
We calculate the radiative power emitted by a blackbody as given by the Stefan-Boltzmann law, $P_{\text{rad}} = \sigma \epsilon T^4$, where $\sigma$ is the Stefan-Boltzmann constant, $\epsilon$ is the emissivity of the surface, and $T$ is the temperature. Using representative refrigerator stage temperatures of $T_{\text{stage}}\in[300.00, 50.00, 4.00, 0.70, 0.10, 0.02]$ K, we estimate the radiative power incident on the MXC resulting from the exposure to the emitter as determined by the area of the LOS apertures. Figure~\ref{fig:passiveload} shows the calculated radiative load as a function of aperture diameter. For a diameter of $d = 10$ cm, corresponding to the geometry used in this work, the radiative link between the $4~$K stage (PT2) and the MXC stage contributes a load of approximately $2.5~$nW, well within the typical cooling power of our dilution fridge ($10~\mu$W at $20~$mK and $400~\mu$W at 100 mK). However,  a LOS to the $50~$K (PT1) stage would produce a much higher burden of $10~\mu$W, whilst the room-temperature stage would contribute a load of $2.3~$mW, exceeding the refrigerator cooling capacities.\\\indent
To put these figures into context, it is useful to compare with the typical passive heat loads produced by a wired set-up used to control a moderate qubit count of 50 (red dashed line), as discussed in~\cite{krinnerEngineeringCryogenicSetups2019}. Although the present work is limited to the wireless excitation of a single resonator, the estimated radiative load associated with the 4 K-to-MXC link remains roughly three orders of magnitude below the wired benchmark. This comparison highlights the potential of wireless interconnects to alleviate thermal-management constraints in future large-scale cryogenic systems.
It is also important to note that the wireless architecture investigated in this work is not intended to establish a direct LOS communication channel between room temperature and the mixing chamber stage. As discussed, such an arrangement would introduce an unsustainable radiative heat load. Instead, the envisioned application of wireless interconnects is communication between cryogenic stages or between cryogenic subsystems already operating within the refrigerator environment, where the thermal burden is substantially reduced. Future implementations involving room-temperature interfaces would require additional infrared filtering, aperture engineering, or staged thermalization to suppress blackbody radiation while preserving microwave transmission.
    
\section{\label{supp:efficiency} Transmitted fraction of radiation}
    
\begin{figure}[t]
        \includegraphics{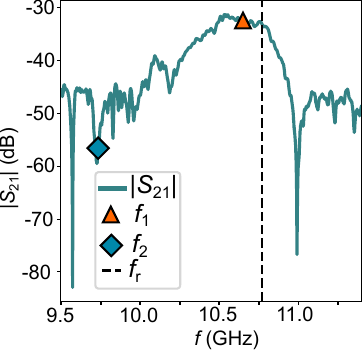}
        \caption{\label{fig:efficiency} Calibrated transmission coefficient as a function of frequency of the stand-alone WLS set-up obtained bypassing the DUT. The values for $f_1$ and $f_2$ are shown as the orange triangle and blue diamond, respectively.}
    \end{figure}

The amount of wireless radiation transferred to the MXC stage was estimated using single-tone measurements generated by an analogue signal source. Figure~\ref{fig:efficiency} highlights two representative frequencies: $f_1 = 10.60$ GHz, located at the center of the metalens passband, and $f_2 = 9.75$ GHz, located outside it. The input powers at the TX was P$_{\text{in}} \in [-13.5\,,\, -10.5\,, -8.5\,]$ dBm after accounting for cable attenuation at 10 GHz. 
    
For each measurement, a continuous-wave signal at either $f_1$ or $f_2$ was applied until the mixing chamber temperature $T_\text{mxc}$ reached a steady value. No microwave signal was collected at the output ports during these measurements; instead, the fraction of broadcast power impinging on the mixing chamber was inferred from the increase in $T_\text{mxc}$ relative to the refrigerator base temperature. The resulting temperature increase was converted to an equivalent absorbed power using a calibration dataset generated by applying a known power to the mixing chamber with the refrigerator's heaters.
    
    \begin{table*}[!htbp] 
        \centering
        \caption{Transmission fractions at $f_1$ and $f_2$}
        \label{tab:efficiency}
        \begin{tabular}{cccccc}
            \toprule
            {$P_{\text{in}}$ (dBm)} & {$P_{\text{in}}$ ($\mu$W)} & {$P_\textup{mxc}(f_1)$ ($\mu$W)} & {$\frac{P_\textup{mxc}}{P_{\text{in}}}(f_1)$} & {$P_\textup{mxc}(f_2)$ ($\mu$W)} & {$\frac{P_\textup{mxc}}{P_{\text{in}}}(f_2)$} \\
            \midrule
            $-13.5$ & $44.67$  & $6.1$  & $13.66\%$ & $3.9$ & $8.73\%$ \\
            $-10.5$ & $89.13$  & $9.0$  & $10.10\%$ & $5.9$ & $6.62\%$ \\
            $-8.5$  & $141.25$ & $11.8$ & $8.35\%$  & $7.9$ & $5.59\%$ \\
            \bottomrule
        \end{tabular}
    \end{table*} 
    
 The results are summarized in Table \ref{tab:efficiency}. We observe a maximum transmission fraction of 13.66\% at $f_1$ for the lowest applied power of -13.5 dBm. At $f_2$, outside the metalens bandwidth, the efficiency is approximately five percentage points lower. As the input power is increased, the power fraction at the mixing chamber stage decreases for both frequencies. Although the exact scaling behavior is not clear, we speculate that either destructive interference may play a more prominent role in the cryostat cavity for higher power intensities or local heating of the Rogers in TX module may introduce a spectral shift away from the intended resonance of the DSRR cell. Nevertheless, these measurements indicate that operation within the metalens passband yields a measurable improvement in transmission fraction with an approximately $60$\% increase relative to frequencies outside the passband region of the spectrum. 

\section{\label{supp:DSRRs}Lens design}
The lens architecture comprises two DSRRs, one at the back and one at the front side. These form a Fabry-Pérot-like cavity that enhances polarization-selective transmission. Each resonator contains two splits defined by an opening angle, and arotation angle to define orientation. A fixed angular separation of 90$^\circ$ is maintained between the splits. By varying the opening and rotation angles of the unit cell, one can achieve independent control of phase delay and arbitrary polarization states. This cross-polarization scheme minimizes radiative crosstalk and suppresses multi-path interference. 

\begin{figure}[t]
    \includegraphics{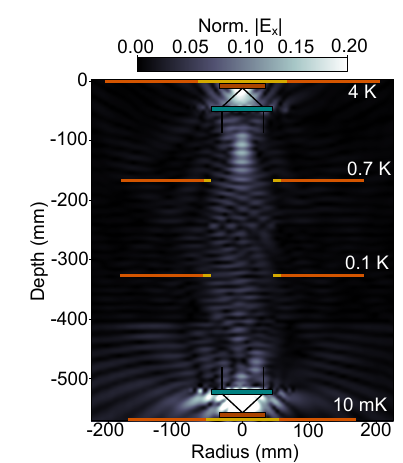}
    \caption{\label{fig:supp_Ex} Simulations of the x-polarized
electric field, $E_\textup{x}$, propagating within the cryostat chamber, complementing the y-polarized signal shown in Fig. \ref{fig:design}(e).
    }
\end{figure}

The metalens was designed with a focal length of F = 50 mm. The required phase profile is given by~\cite{al-moathinCharacterizationCompactWideband2023}, 
\begin{equation}
        \Delta\varphi = \frac{2\pi}{\lambda}(\sqrt{F^2 + x^2 + y^2} - F),
\end{equation}

where $\lambda$ is the wavelength and $x$ and $y$ denote the positions relative to the lens center. To account for the spherical wavefront produced by the antenna, an additional phase correction was incorporated such that the target phase shift was controlled through geometric parameters of each unit cell (e.g. opening and rotation angles).\\\indent
Figure~\ref{fig:supp_Ex} shows numerical simulations of the x-polarized electric field propagation within the cryostat, which complement  the y-polarized simulations of Fig.~\ref{fig:design}(e). The patch antenna transmits the linear x-polarized RF signal that is rotated to $ E_\text{y} $ by the DSRR metalenses. The data shows a significant reduction in the $|E_{\text{x}}|$ electric field intensity which propagates across the apertures with respect to $|E_{\text{y}}|$. Yet, collimated residual x-polarized light is present, as shown in Fig.~\ref{fig:supp_Ex}.  The effect of the RF absorbers introduced to attenuate the signal that is not captured by the DSRRs has not been simulated. 
\section{\label{supp:resonator} Superconducting resonator model}
Evaluating resonator parameters from the frequency-dependent transmission response in a notch geometry is typically modeled~\cite{probstEfficientRobustAnalysis2015a, mcraeMaterialsLossMeasurements2020} by, 
    \begin{equation}\label{eqn:notch_res}
        |S_{21}|(f) = ae^{i\alpha}e^{-2\pi i f \tau}[1 - \frac{(Q_L / |Q_c|)e^{i\phi}}{1 + 2iQ_L(f/f_r - 1)}].
    \end{equation}
Here, $a$, $\alpha$, and $\tau$ account for amplitude scaling, phase offset, and electrical delay, respectively, while $\phi$ captures impedance mismatch. The resonator parameters, including loaded, internal, and coupling quality factors, can then be extracted using the circle-fitting method of Probst \textit{et al.}~\cite{probstEfficientRobustAnalysis2015a}.\\\indent
The calculated $\Delta f \approx \Delta f_{\text{qp}}$ line of Fig.~\ref{fig:CPW}(c) was evaluated using the expression~\cite{foshatQuasiparticleDynamicsNbN2025}, 
    \begin{equation}\label{eqn:delta_qp}
        \Delta f_{\text{qp}} = -\frac{1}{2}\alpha_{L,ki} f_r \frac{\Delta}{k_B T \sinh(\Delta/k_B T)},
    \end{equation}
where $\alpha_{L,ki} = 0.974$ is the kinetic inductance ratio evaluated from previous measurements of the resonators in~\cite{foshatCharacterizingNiobiumNitride2023}, and $\Delta$ is the superconducting gap.\\ 
\section{\label{supp:Q3D}Enclosure parasitic capacitance}
\begin{figure}[b]
        \centering
        \includegraphics[width=0.7\columnwidth]{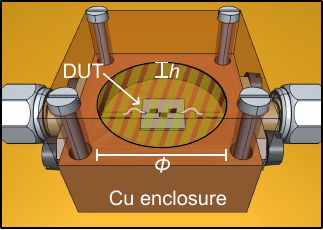}
         \caption{\label{fig:suppQ3D}
         Rendering of the 3D geometry used to estimate the parasitic capacitance between the DUT and its copper enclosure. $C_\text{p}$ is evaluated as the capacitance between the on-chip feedline and the walls of the recess (shaded yellow) of height $h = 3.5~$mm and diameter  $\phi = 17~$mm.}
    \end{figure}

Electrostatic simulations in \textit{Ansys Q3D Extractor} enable an estimate of the parasitic capacitance, $C_\textup{p}$, which contributes to the parasitic, $Z_\textup{p}$, discussed in Section \ref{sec:parasitic_wless}. Stray wireless signals are expected to form spurious modes within the copper enclosure through which the device can couple. Figure \ref{fig:suppQ3D} displays the simulated structure, including the NbN resonator, silicon substrate, coplanar waveguide feedline, and the surrounding copper enclosure. The yellow shaded cylinder (height $h$ = 3.5 mm, diameter $\phi = 17$ mm) shows the internal cavity where $C_\text{p}$ is extracted from the simulation. This allows for the extraction of the quasi-static geometric capacitance, but it does not involve full microwave loss simulations.\\\indent 
The extracted capacitance matrix found the mutual capacitance between the on-chip feedline (that forms part of the readout line) and the copper enclosure to be $C_\text{p} \simeq 40$ fF. This is roughly on par with the resonator-feedline coupling capacitance, $C_{\text{feed}} \sim$ 15-25 fF, which was evaluated separately using information of the device coupling geometry. This corroborates the argument that the additional stray coupling path may contribute significantly to the loaded DUT quality factor, as discussed in Section \ref{sec:parasitic_wless}.

\bibliography{ref_kb}

\end{document}